\documentclass[a4paper,fleqn,usenatbib]{mnras}

\usepackage{amsopn}
\DeclareMathOperator{\erf}{erf}
\usepackage{color}
\usepackage{caption}
\usepackage{relsize}
\usepackage{graphicx}
\usepackage{amsmath}
\usepackage{amssymb}
\usepackage{pdflscape}
\usepackage{rotating}
\usepackage{longtable}
\usepackage{afterpage}

\title[Bimodal radio variability in blazars]{Bimodal radio variability in OVRO-40m-monitored blazars}
\author[Liodakis et al.]
{I. Liodakis$^{1,2}$\thanks{liodakis@physics.uoc.gr}, V. Pavlidou$^{1,2}$, T. Hovatta$^{3,4,5}$, W. Max-Moerbeck$^{6}$, T. J. Pearson$^{7}$,\newauthor J. L. Richards$^{7}$, and  A. C. S. Readhead$^{7}$ \\
$^{1}$Department of Physics and ITCP\thanks{Institute for Theoretical and Computational Physics, formerly Institute for Plasma Physics}, University of Crete, 71003, Heraklion, Greece\\
$^{2}$Foundation for Research and Technology - Hellas, IESL, Voutes, 7110 Heraklion, Greece\\
$^{3}$Aalto University Mets\"ahovi Radio Observatory, Mets\"ahovintie 114, 02540 Kylm\"al\"a, Finland\\
 $^{4}$Aalto University Department of Radio Science and Engineering,P.O. BOX 13000, FI-00076 AALTO, Finland\\
$^{5}$Tuorla Observatory, Department of Physics and Astronomy, University of Turku, Finland\\
$^{6}$Max-Planck-Institut f\"ur Radioastronomie, Auf dem H\"ugel 69, 53121 Bonn, Germany \\
$^{7}$Cahill Center for Astronomy and Astrophysics, California Institute of Technology, 1200 E California Blvd, Pasadena, CA 91125, USA\\
}

\begin{document}
\maketitle
\label{firstpage}
\begin{abstract}
Blazars are known to show periods of quiescence followed by outbursts visible throughout the electromagnetic spectrum. We present a novel maximum likelihood approach to capture this bimodal behavior by examining blazar radio variability in the flux-density domain. We separate quiescent and flaring components of a source's light curve by modeling its flux-density distribution as a series of ``off'' and ``on'' states. Our modeling allows us to extract information regarding the flaring ratio, duty cycle, and the modulation index in the ``off''-state, in the ``on''-state, as well as throughout the monitoring period of each blazar. We apply our method to a flux-density-limited subsample from the Owens Valley Radio observatory's 15 GHz blazar monitoring program, and explore differences in the variability characteristics between BL Lacs and FSRQs as well as between $\gamma$-ray detected and non-detected sources. We find that: (1) BL Lacs are more variable and have relatively larger outbursts than the FSRQs; (2) unclassified blazar candidates in our sample show similar variability characteristics as the FSRQs; and (3) $\gamma$-ray detected differ from the $\gamma$-ray non-detected sources in all their variability properties, suggesting a link between the production of $\gamma$-rays and the mechanism responsible for the radio variability. Finally, we fit distributions for blazar flaring ratios, duty cycles, and on- and off- modulation indices that can be used in population studies of variability-dependent blazar properties.
\end{abstract}


\section{Introduction}\label{introduc}
BL Lac objects (BL Lacs) and Flat Spectrum Radio Quasars (FSRQs) constitute a sub-class of active galactic nuclei (AGN) called blazars. Blazars are known for their powerful and highly relativistic jets, which are pointed close to our line of sight \citep{Readhead1978,Blandford1979,Scheuer1979,Readhead1980}. Due to the alignment of the jet, their emission, from radio to the highest energy $\gamma$-rays, is dominated by relativistic effects such as boosting of the observed flux and compression of timescales. They show a compact one-sided core-dominated jet morphology, as well as apparent superluminal motion of the radio jet components propagating downstream of the core as seen through VLBI observations \citep{Readhead1978,Scheuer1979,Readhead1980}. Such radio components have shown superluminal motions of up to about $50c$ \citep{Lister2009,Lister2013}, while they have also been associated with numerous other phenomena such as the production of $\gamma$-rays and rotations of the optical polarization plane \citep{Marscher2008,Marscher2010}. 

Variability in the radio regime was one of the first identified characteristics of blazars \citep{Dent1965}. Although there are cases of quasi-periodicity \citep{Carrasco1985,Valtaoja1985,King2013}, generally the variability is erratic. However, it cannot be fully explained by a stochastic process, due to the appearance of outbursts, often in several frequency bands simultaneously \citep{Aller1999,Hovatta2008,Max-Moerbeck2014,Fuhrmann2014,Blinov2015,Angelakis2016,Hovatta2016}, followed by periods of relatively low activity. It has been suggested that such outbursts can be characterized by an exponential rise and decay with the ratio of the respective timescales to be approximately 1.3 \citep{Valtaoja1999,Lahteenmaki1999-II,Lahteenmaki1999-III,Hovatta2009}. The timescale and flux-density amplitude of an outburst can vary from days to months and from comparable, to orders of magnitude higher than the quiescent one \citep{Aller1999}. Thus the radio flux-density curve of a blazar can generally be described by a quiescent level (the minimum radio output of the source), and a series of consecutive aperiodic outbursts on top of that minimum output.

Modeling the variability of blazars has been the subject of several studies. \cite{Valtaoja1988} attempted to separate the quiescent from the flaring flux through multi-wavelength flux-density curves (several frequencies from 4.8-90~GHz). The authors examined the spectrum of each source at periods of minimum flux between outbursts (what they considered as the ``constant'' flux of the jet), and subtracted it from the source spectrum during an outburst in order to obtain the ``variable'' flux. \cite{Lister2001} chose to model the blazar flux-density curves (4.8 and 5~GHz) with a shot-noise process (\citealp{Lister2001}, and references therein). Using Poisson statistics [and the exponential profile for the flares described in  \cite{Valtaoja1999,Lahteenmaki1999-II,Lahteenmaki1999-III}] the authors investigated the variability duty cycles and sample selection biases due to variability in blazars.

The interest in the subject is well-motivated: understanding the general variability properties of blazars in different frequencies can provide important information on their emission mechanisms as well as the location of the emission region. Although rest-frame time delays between frequencies are necessary to constrain the location of emission regions on a source-by-source basis \citep{Fuhrmann2014,Max-Moerbeck2014}, differences in the variability properties at different frequency bands could be indicative of the spatial connection of their respective emission regions on a population level.

The OVRO 40m monitoring program \citep{Richards2011,Richards2014} provides a unique opportunity for studies of blazar radio variability, thanks to its unprecedented sample size and cadence (about 1800 sources observed twice weekly on average over 8 years). OVRO data from  2 years of monitoring were used in \cite{Richards2011} to examine blazar variability and amplitudes using a likelihood approach. Modeling each source's flux-density distribution by a single Gaussian distribution, the ``intrinsic'' mean flux-density and modulation indices (what one would have observed in the limit of infinite accuracy and sampling) were estimated. The authors used these results to uncover a statistically significant discrepancy between the radio variability (as quantified by the intrinsic modulation index) of $\gamma$-ray loud and $\gamma$-ray quiet sources that were otherwise similar. For this reason, the radio intrinsic modulation index has since been used to select $\gamma$-ray quiet sources that are as similar in their radio properties as possible to $\gamma$-ray loud blazars, for monitoring in other frequencies (e.g. \citealp{Pavlidou2014,Angelakis2016}). 

The likelihood formalism used by \cite{Richards2011} provides a robust way to account for observational uncertainties and finite cadence as well as calculate uncertainties for the estimated quantities, but at the expense of model-independence. In this case, model-dependence enters through the assumption that a single-Gaussian is a good description of the distribution of flux-densities. If that assumption is not valid for some sources, this can affect the results in two ways. First, any estimated quantities that are sensitive to the assumed underlying family of flux-density distributions may exhibit systematic offsets. Most importantly, these offsets could {\it not} be accounted for by the calculated uncertainties, which are statistical and based on the assumption that the underlying distribution model holds (see e.g., \citealp{Mouschovias2010}). Second, important information regarding the detailed behavior of the source (such as bimodality in the flux-density distribution) is lost through the simplified treatment.

As discussed in \cite{Richards2011}, some of the OVRO 40m-monitored blazars show a flux-density distribution that can be well described by a single Gaussian, while others show a bimodal distribution. To remedy the first of the two problems described above (i.e. cases where a single Gaussian model is not a good description of the flux-density distribution) \cite{Richards2011} used, and reported, only quantities that their method estimates robustly. Such quantities are the mean flux and the modulation index, in contrast, for example with the most likely flux of a source. A typical example of such a source is BL Lac, the flux distribution of which (using the full 8 years OVRO dataset) is shown in Fig. \ref{plt:prototype}. The maximum-likelihood single Gaussian for this data set is overplotted with the dashed line. As BL Lac is very bright, and the observational uncertainties are very small compared to the typical flux ($<2\%$), we do not expect the uncertainties to widen the flux distribution significantly, or, conversely, the maximum-likelihood distribution to be appreciably narrower than the data. Indeed, this is the case in Fig. \ref{plt:prototype}. The source exhibits bimodality. For this reason, the maximum-likelihood single Gaussian is {\em not} a good description of the underlying flux-density distribution. As a result, while the single-Gaussian mean and spread are clearly reasonable representations of the corresponding parameters of the data distribution, the most-likely flux is missed by the single Gaussian, and information such as the flaring ratio (ratio of the mean ``on''-state flux-density to ``off''-state flux density $R=S_{\rm on}/S_{\rm off}$) is lost. On the other hand, a model that can account for the bimodality of the sources can recover this important information. Such a model is particularly important given the large size, good sampling, low noise and statistical completeness of the dataset at hand.

In this work, we extend the formalism of \cite{Richards2011} in order to model the observed blazar variability as an alternation between a low-flux state and a high-flux state with the possibility of different degrees of variations about the mean flux of each state. The physical interpretation of such a model is that blazars spend a fraction of their time in a state characterized by relatively low or no activity, and the remaining time in a state of increased variability and outbursts. This way we attempt to separate the quiescent and flaring components of the flux-density distribution of each source, and examine the general variability characteristics, as well as differences between different blazar subclasses. The paper is organized as follows. In section \ref{MLE} we present the maximum likelihood formalism we developed in order to model the blazar flux-density distributions. In section \ref{OVRO_samp} we present the sample used in this work. In section \ref{single_vs_bimodal} we compare the results from our model with that of a single Gaussian distribution and test the bimodality of the flux-density distributions of the sources in our sample. In section \ref{Var_analysis} we derive best-fit distributions of all quantities characterizing bimodal variability for the blazar population and we compare the variability characteristics of different blazar subsamples. Finally in section \ref{discussion} we summarize and discuss our findings and conclusions.

The cosmological parameters we adopt in this work are $H_0=71$ ${\rm km \, s^{-1} \, Mpc^{-1}}$, $\Omega_m=0.27$ and $\Omega_\Lambda=1-\Omega_m$ \citep{Komatsu2009}.

\section{Maximum likelihood analysis}\label{MLE}
\begin{figure}
\resizebox{\hsize}{!}{\includegraphics[scale=1]{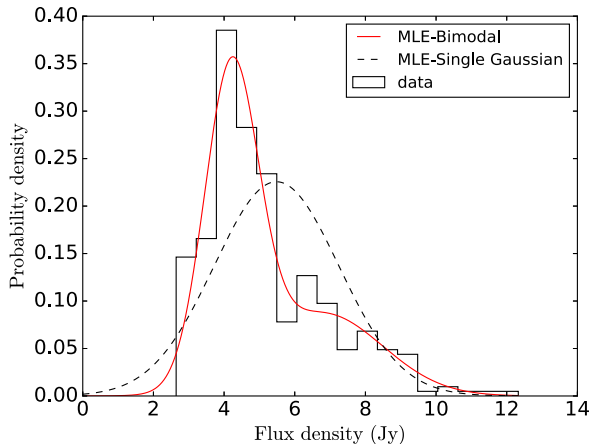} }
 \caption{Distribution of the 15 GHz flux-density of BL Lac. The red line represents the maximum likelihood fit of the bimodal model whereas the dashed black is the fit for the single Gaussian model \citep{Richards2011}.}
 \label{plt:prototype}
 \end{figure}
\begin{figure}
\resizebox{\hsize}{!}{\includegraphics[scale=1]{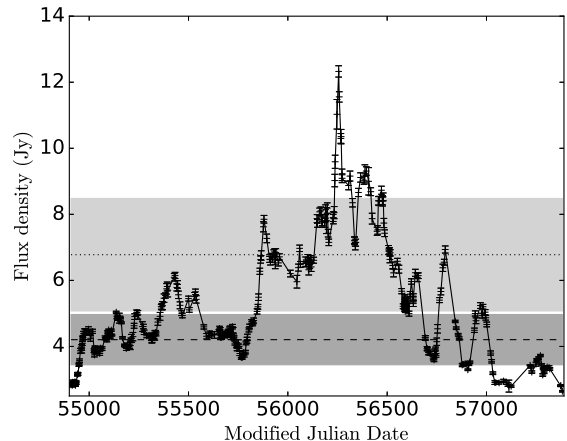} }
\caption{OVRO radio (15~GHz) light curve of BL Lac. The dashed line marks the mean  flux-density of the ``off''-state while the dotted line the mean flux-density of the ``on''-state. The dark grey and light grey shaded areas are the 1-standard deviation of the ``off-state'' and ``on''-state flux-densities respectively.}
\label{plt:lightcurve_bllac}
\end{figure} 
At any time of observation the flux-density ($S$) emitted by a source, given the observational uncertainty (which we assume to be Gaussian), can be described as
\begin{eqnarray}
P(S|S_{\rm obs},\sigma_{\rm obs})=\frac{1}{\sigma_{\rm obs}\sqrt{2\pi}}\exp\left[-\frac{(S-S_{\rm obs})^2}{2\sigma_{\rm obs}^2}\right],
\end{eqnarray}
where $S_{\rm obs}$ is the observed flux-density, and $\sigma_{\rm obs}$ is the error on the measurement. We model the flux-density distribution of blazars as a sequence of two states dubbed ``off'' and ``on''. Each state can be assumed to follow a Gaussian distribution similarly to \cite{Richards2011}. The ``off''-state describes the period of time a blazar spends in quiescence, while the ``on''-state the period of time a blazar spends flaring. Then the probability density function of the blazar emitting a certain flux-density will be given by,
\begin{eqnarray}
P(S|\mathsmaller{\mathsmaller{f_{\rm t},S_{\rm off},S_{\rm on},\sigma_{\rm off},\sigma_{\rm off}}})=\mathsmaller{\frac{1-f_{\rm t}}{\sigma_{\rm off}\sqrt{2\pi}}\exp\left[-\frac{(S-S_{\rm off})^2}{2\sigma_{\rm off}^2}\right]}\nonumber\\
+\mathsmaller{\frac{f_{\rm t}}{\sigma_{\rm on}\sqrt{2\pi}}\exp\left[-\frac{(S-S_{\rm on})^2}{2\sigma_{\rm on}^2}\right]},
\end{eqnarray}
where $f_{\rm t}$ is the duty cycle, $S_{\rm off}$ is the mean flux-density of the ``off''-state, $\sigma_{\rm off}$ is the standard deviation of the mean flux-density of the ``off''-state, $S_{\rm on}$ is the mean flux-density of the ``on''-state, and $\sigma_{\rm on}$ is the standard deviation of the mean flux-density of the ``on''-state.

Marginalizing over all possible flux-densities S (limits of the integral from $-\infty$ to $\infty$, see \cite{Venters2007,Richards2011} for a detailed derivation of the integral), the likelihood of observing $f_{\rm t}$, $S_{\rm off}$, $S_{\rm on}$, $\sigma_{\rm off}$, and $\sigma_{\rm on}$ given a flux-density $S_{\rm obs}$ and error $\sigma_{\rm obs}$ is,
 \begin{eqnarray}\label{eq:likelihood}
l_{\rm obs}&=&\frac{1-f_{\rm t}}{\sqrt{2\pi(\sigma_{\rm off}^2+\sigma_{\rm obs}^2)}}\exp\left[-\frac{(S_{\rm off}-S_{\rm obs})^2}{2(\sigma_{\rm off}^2+\sigma_{\rm obs}^2)}\right]\nonumber\\
&+&\frac{f_{\rm t}}{\sqrt{2\pi(\sigma_{\rm on}^2+\sigma_{\rm obs}^2)}}\exp\left[-\frac{(S_{\rm on}-S_{\rm obs})^2}{2(\sigma_{\rm on}^2+\sigma_{\rm obs}^2)}\right].
\end{eqnarray}
The first term of Eq. \ref{eq:likelihood} describes the ``off''-state, while the second the ``on''-state. For a number of observations $N$ ($j=1,2,3...N$),  the likelihood function is,
\begin{eqnarray}
\mathcal{L}=\prod_{j=1}^N l_{\rm obs,j}\Rightarrow \log\mathcal{L}=\sum_{j=1}^N \log l_{\rm obs,j}.
\end{eqnarray}
We use the Simplex algorithm (also known as Nelder-Mead, \citealp{Nelder1965}) implemented in the scipy.optimize.minimize python package to minimize the negative log-likelihood, and thus to maximize the likelihood. After several trials with different methods and minimizing routines implemented in various programming environments, the Simplex algorithm was proven to be the most consistent and stable, given the task at hand.

Given the complexity of the likelihood function, in order to avoid numerical instabilities and possible sensitivity to initial conditions, we perform the minimization $10^3$ times, each time with random initial conditions, and choose as the best-fit parameter estimates those that gave the minimum function value. Figure \ref{plt:prototype} shows the maximum likelihood fit of our method to the 15 GHz flux-density distribution of BL Lacertae  while Fig. \ref{plt:lightcurve_bllac} shows the OVRO light curve. The horizontal lines mark the mean flux-densities of the ``off''- and ``on''-states and the grey shaded areas their 1-standard deviation.

To derive the error on our estimated parameter values, we use the Fisher information matrix. The information matrix gives a measure of the amount of information every parameter carries on the curvature of the likelihood at the best-fit values. From the information matrix, we calculate the variance-covariance matrix. The error on each parameter estimate is the square root of the corresponding diagonal element of the variance-covariance matrix. It is possible that the matrix is not positive definite. Since the likelihood is multi-parametric, this would suggest that two or more parameters are anti-correlated. In such a case, we estimate the error of that parameter using a slice of the likelihood surface parallel to the axis of the parameter of interest, with the values of the other parameters set at their best-fit values (i.e., a slice passing through the maximum likelihood point). Using that slice, we determine the values of the parameters that reduce the likelihood by factor $e^{-1/2}$ (in a Gaussian slice, this would be the $\pm 1\sigma$ points). These two values set the 1$\sigma$ uncertainty on that parameter.

The method described above can provide robust results as long as the number of observations is sufficiently large. In the application of our method (see section \ref{OVRO_samp}) the number of observations per source is 421 on average with a standard deviation of 89 (minimum 217 and maximum 1108).

\section{Sample}\label{OVRO_samp}
\begin{figure}
\resizebox{\hsize}{!}{\includegraphics[scale=1]{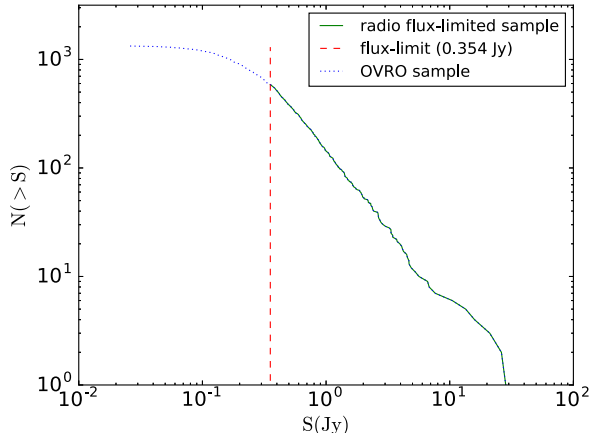} }
 \caption{Source count distribution for the OVRO sources. The red dashed line marks the flux limit at 0.354 Jy. }
 \label{plt:flux_limit}
 \end{figure}

We apply our method to data from the Owens Valley Radio Observatory's (OVRO) 15 GHz blazar monitoring program\footnote{http://www.astro.caltech.edu/ovroblazars/} \citep{Richards2011}. OVRO has been monitoring a sample of $>$1800 blazars in support of the {\it Fermi} Gamma-ray Space Telescope ({\it Fermi}, \citealp{Acero2015}) with an approximate cadence of twice per week since 2007. The monitoring program began using the 1158 sources from the Candidate Gamma-ray Blazar Survey (CGRaBS) complete sample \citep{Healey2008}. Since then, however, to facilitate monitoring of additional sources detected by {\it Fermi}, the sample has increased by approximately 1/2 of its original size. In order to maintain the statistical integrity on the sample under investigation while taking advantage of the additional $\gamma$-ray sources that were added later, we construct a new statistically complete sample (a flux-density--limited one), as follows: we use the maximum likelihood mean flux-density from \cite{Richards2014} to plot the source count distribution for all OVRO monitored sources (Fig. \ref{plt:flux_limit}). The flux limit is set at 0.354 Jy which is where the distribution shape starts to deviate from a power law. The final sample consists of 584 sources; 435 FSRQs, 81 BL Lacs, and 68 other sources (which we dubbed U-R), 17 of which are classified as radio galaxies and 51 as possible blazar or unidentified. We follow the classification of \cite{Richards2011,Richards2014} (and references therein) i.e. objects with broad emission lines in optical are classified as FSRQs whereas objects with weak or even no emission lines are classified as BL Lacs. We have included data from 05/01/2008 to 14/02/2016.

\section{Single Gaussian versus bimodal}\label{single_vs_bimodal}
\begin{table}
\setlength{\tabcolsep}{12pt}
\centering
  \caption{Mean parameter values and probabilities for the Wilcoxon and K-S test that the $S$ and $m$ parameters estimated by the two models (single-Gaussian and bimodal) are consistent (drawn from the same distribution).}
  \label{tab:1g_vs_2g}
\begin{tabular}{@{}ccccc@{}}
 \hline
    Parameter  & Single & Bimodal & Wilcoxon & K-S \\
   & (mean)  & (mean)  & (\%) & (\%)  \\
  \hline
    $S$ & 1.11  & 1.11 & 90.0  & 99.5  \\
    $m$ & 0.18 & 0.20 & 0.28 & 1.26  \\
\hline
\end{tabular}
\end{table}
\begin{figure}
\resizebox{\hsize}{!}{\includegraphics[scale=1]{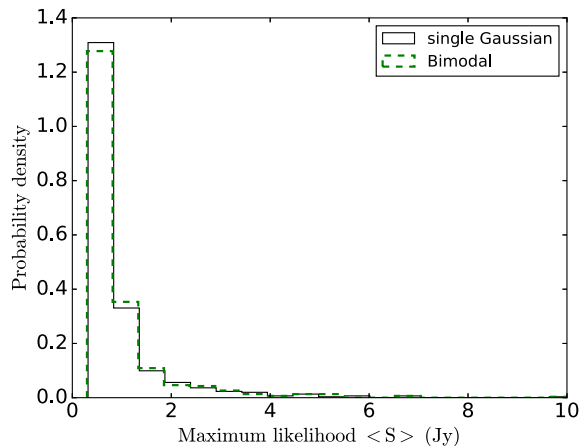} }
 \caption{Distribution of the maximum likelihood $\mathrm{\langle S \rangle}$ for the single Gaussian and bimodal models.}
 \label{plt:comp_flux}
 \end{figure}
\begin{figure}
\resizebox{\hsize}{!}{\includegraphics[scale=1]{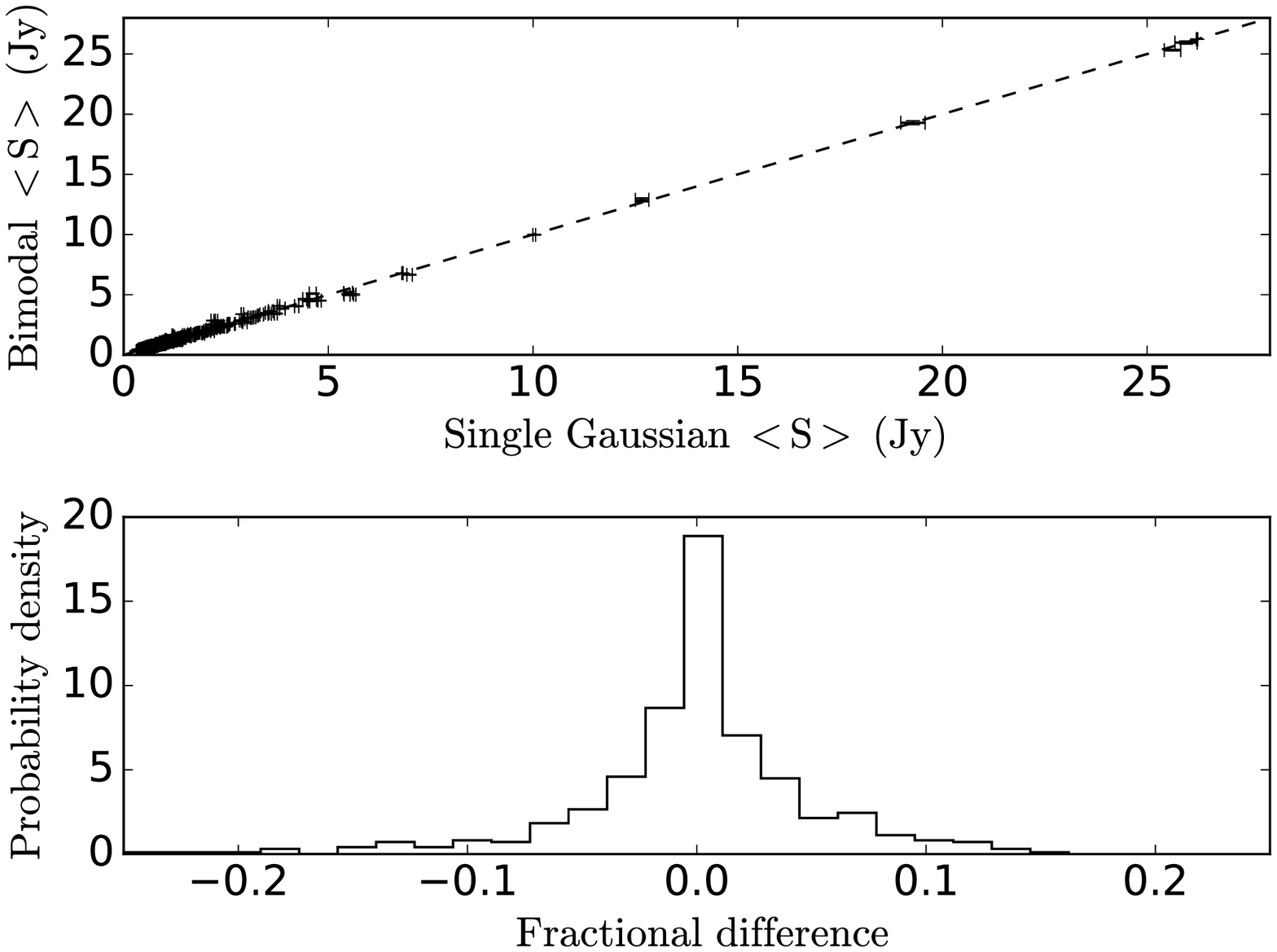} }
 \caption{{\bf Upper panel:} Bimodal MLE $\mathrm{\langle S\rangle}$ versus single Gaussian MLE $\mathrm{\langle S\rangle}$. The dashed line corresponds to $y=x$. {\bf Lower panel:} Distribution of the fractional difference between the bimodal $\mathrm{\langle S\rangle}$ and single Gaussian $\mathrm{\langle S\rangle}$. }
 \label{plt:diff_comp_flux}
 \end{figure} 

\begin{figure}
\resizebox{\hsize}{!}{\includegraphics[scale=1]{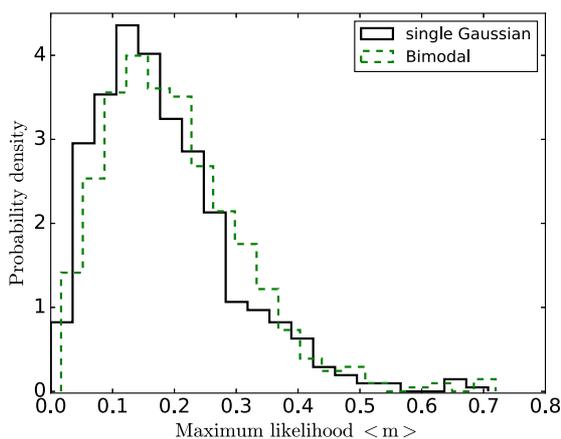} }
 \caption{Distribution of the intrinsic modulation index for the single Gaussian and bimodal models.}
 \label{plt:comp_mod}
 \end{figure}
\begin{figure}
\resizebox{\hsize}{!}{\includegraphics[scale=1]{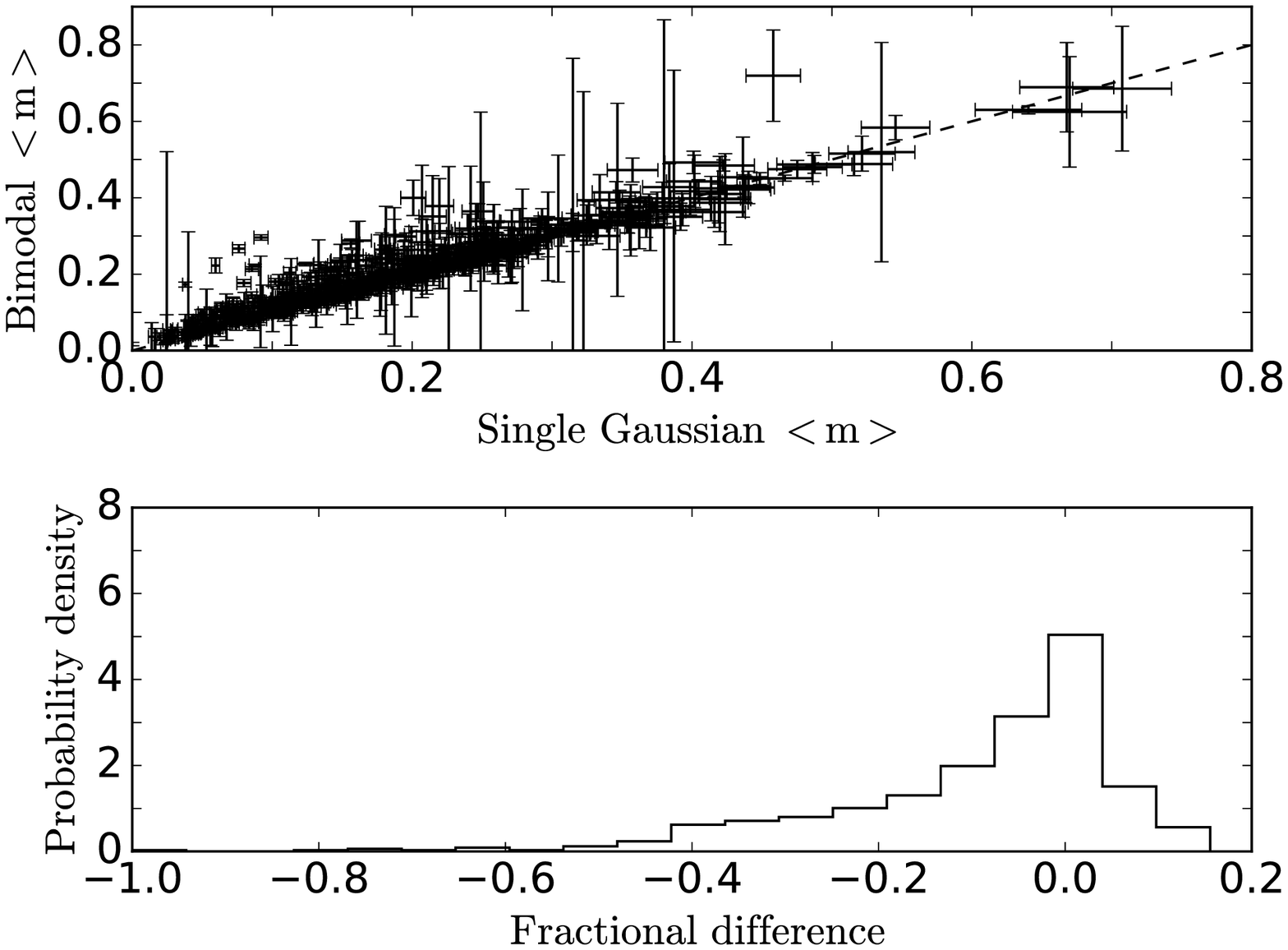} }
 \caption{{{\bf Upper panel:} Bimodal MLE $\mathrm{\langle m\rangle}$ versus single Gaussian MLE $\mathrm{\langle m\rangle}$. The dashed line corresponds to $y=x$. {\bf Lower panel:} Distribution of the fractional difference between the bimodal MLE $\mathrm{\langle m\rangle}$ and single Gaussian MLE $\mathrm{\langle m\rangle}$. }}
 \label{plt:diff_comp_mod}
 \end{figure}  
\begin{figure}
\resizebox{\hsize}{!}{\includegraphics[scale=1]{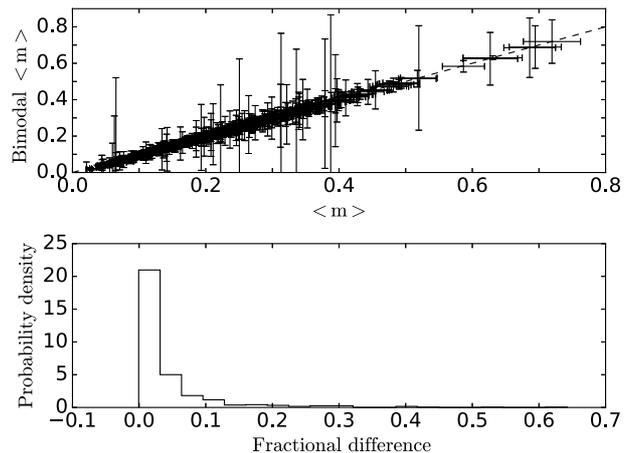} }
 \caption{{{\bf Upper panel:} Bimodal likelihood $\mathrm{\langle m\rangle}$ versus sample $\mathrm{\langle m\rangle}$. The dashed line corresponds to $y=x$. {\bf Lower panel:} Distribution of the fractional difference between the bimodal MLE $\mathrm{\langle m\rangle}$ and sample $\mathrm{\langle m\rangle}$. }}
 \label{plt:diff_comp_mod_nl}
 \end{figure}    
\begin{figure}
\resizebox{\hsize}{!}{\includegraphics[scale=1]{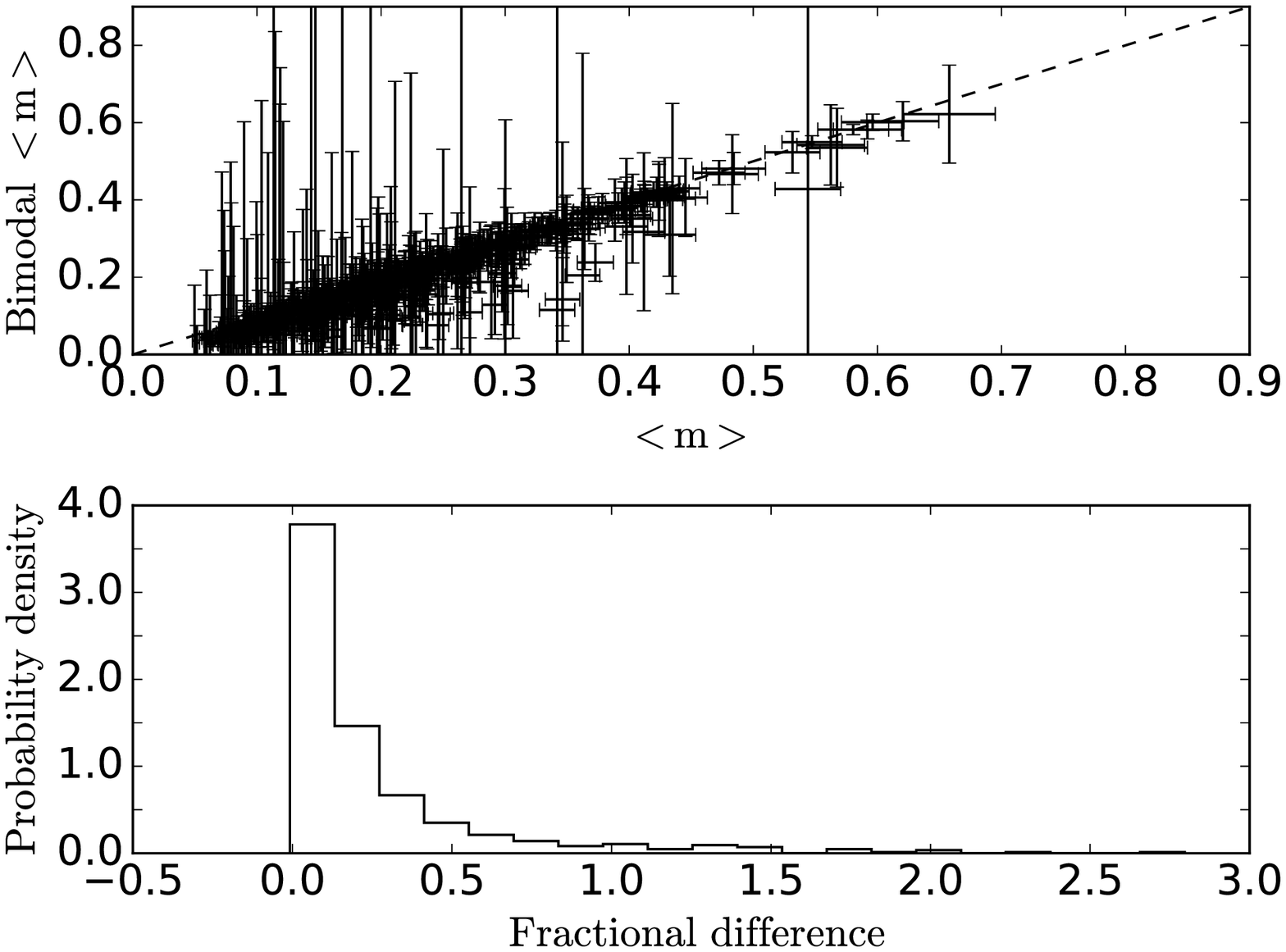} }
 \caption{{{\bf Upper panel:} Bimodal MLE $\mathrm{\langle m\rangle}$ versus sample $\mathrm{\langle m\rangle}$ for the sources below the flux-density limit. The dashed line corresponds to  $y=x$. {\bf Lower panel:} Distribution of the fractional difference between the bimodal MLE $\mathrm{\langle m\rangle}$ and sample $\mathrm{\langle m\rangle}$  for the sources below the flux-density limit.}}
 \label{plt:diff_comp_mod_noflux-limit}
 \end{figure} 

Using the method described in section \ref{MLE} we fit the flux-density distribution of every source in the flux-limited subsample defined above, in order to extract their variability properties. It is not necessary that the flux-density distribution of a source is described well by a ``off''-``on'' model. It is very well possible that a source resides in either state for the entire duration of the monitoring period, which is roughly 8 years. In this case, the maximum likelihood best-fit model is degenerate. Models with duty cycle $f_{\rm t}=1$, $f_{\rm t}=0$, or two identical states with $f_{\rm t}=0.5$ are mathematically possible and describe the same single Gaussian distribution. For this reason, in this section we compare the results of our model and the single Gaussian model as in \cite{Richards2011} (fit to the 8-year dataset) in order to understand which model best describes the data, and identify all sources with such degeneracy. For the purposes of our analysis we use the Kolmogorov-Smirnov test (K-S test) and the Wilcoxon rank-sum test (Wilcoxon test). The K-S test gives the probability of two samples being drawn from the same distribution, while the Wilcoxon test gives the same probability but with the alternative hypothesis that the values of one sample are systematically larger than those of the other. For any probability higher than 5\% we accept that neither test can reject the null hypothesis that the two samples are drawn from the same distribution. Only 52 sources ($8.9\%$) can be well-described by a single-Gaussian model. For the purposes of the population studies (see section \ref{Var_analysis}) we have excluded these sources from our analysis. The excluded sources also account for cases where the source has no significant variability, since the flux-density distribution of such a source would be consistent with a single narrow Gaussian distribution.

We next proceed to a comparison between the overall mean flux-density and modulation index (in contrast to the mean and modulation index of the individual variability states) of the bimodal model with the corresponding qualities of the single-Gaussian model, in order to evaluate any systematic effects on these quantities due to the single-Gaussian assumption.

From the definition of the mean and variance, the mean flux-density of the bimodal distribution is equal to,
\begin{equation}
\langle S\rangle=(1-f_{\rm t}) S_{\rm off} + f_t S_{\rm on},
\label{eq:over_flux}
\end{equation} 
and the variance is
\begin{equation}
{\rm Var}= [(1-f_{\rm t})(\sigma_{\rm off}^2 + S_{\rm off}^2) + f_t (\sigma_{\rm on}^2+S_{\rm on}^2)]-\langle S\rangle^2.
\label{eq:over_var}
\end{equation}
Using Eq. \ref{eq:over_flux} and \ref{eq:over_var}  the overall modulation index is equal to $\mathrm{\langle m\rangle=\sqrt{\rm Var}/\langle S\rangle}$. We calculate the overall mean flux-density and intrinsic modulation index for each of our sources, and compare it with the corresponding values from a single Gaussian model from \cite{Richards2011}. In order to estimate uncertainties on $\mathrm{\langle S\rangle}$ and $\mathrm{\langle m\rangle}$ we use the uncertainties estimated by the likelihood analysis for ${\rm f_t}$, $S_{\rm on}$, $S_{\rm off}$, $\sigma_{\rm on}$, and $\sigma_{\rm off}$, and standard error propagation.

For the mean flux-density the mean of the two samples is the same with both tests, while both K-S and Wilcoxon tests allow for the hypothesis that the samples are drawn from the same distribution (Table \ref{tab:1g_vs_2g}). Figure \ref{plt:comp_flux} shows the distribution of the overall $\mathrm{\langle S\rangle}$ for the single Gaussian (solid black line) and bimodal (dashed green line) models and Fig. \ref{plt:diff_comp_flux} shows the comparison of the two models for individual sources (upper panel) and the distribution of the fractional difference $\mathrm{(\langle S_{\rm Gaussian}\rangle-\langle S_{\rm Bimodal}\rangle)/\langle S_{\rm Bimodal}\rangle}$, between models (lower panel). It is clear that the mean is very robust against the single-Gaussian assumption. The distribution of fractional differences is symmetric about zero, indicating that there is no bias in the mean introduced by the single-Gaussian assumption.

For the modulation index, Fig. \ref{plt:comp_mod} shows the distribution of the overall $\mathrm{\langle m\rangle}$ for both models and Fig. \ref{plt:diff_comp_mod} shows the comparison of the two models for individual sources (upper panel) and the distribution of their fractional difference (lower panel). Both K-S and Wilcoxon test indicate disagreement between the two estimates, although not at extremely high significance (Table \ref{tab:1g_vs_2g}). The scatter between models is larger than with the $\mathrm{\langle S\rangle}$ (Fig. \ref{plt:diff_comp_mod}). However, for most sources the values are consistent within uncertainties, and the fractional difference (lower panel of Fig. \ref{plt:diff_comp_mod}) is generally less than 20\% with the exception of very few sources. For this reason, we do not expect the single-Gaussian assumption to have had a strong impact on the statistical comparison between populations in \cite{Richards2011} and \cite{Richards2014}. It is nevertheless worth noting that the single-Gaussian assumption introduces a negative bias in the modulation index: the distribution of fractional difference between models is not symmetric about zero (mean$\approx$-0.09, median$\approx$-0.04) with the single-Gaussian model tending to underestimate the modulation index and the overall variability of the sources.

The discrepancy originates from the fact that, for a source well-described by a bimodal distribution (e.g. Fig. \ref{plt:prototype}), a single Gaussian model would favor the dominant peak (in that case the ``off''-state) and attempt to accommodate the alternate state in the tail of the distribution. That is why although the $\mathrm{\langle S\rangle}$ derived by the two models are consistent, the $\mathrm{\langle m\rangle}$ are not.

On the other hand, if we compare the overall $\mathrm{\langle m\rangle}$ from the bimodal model (maximum-likelihood estimate, MLE) with a standard sample $\mathrm{\langle m\rangle}$ (ratio between sample standard deviation and sample mean, Fig. \ref{plt:diff_comp_mod_nl}) we find that they are in good agreement within uncertainties. The Wilcoxon test yields a $45\%$ and the K-S test a $87\%$  probability of consistency respectively. In order to arrive to such result, two conditions need to be met. First, the model we have assumed (bimodal) has to be a good description of the underlying distribution of the sources, and second the observational errors have to be relatively small compared to the intrinsic variability of the sources. The latter is expected for the sources in our sample, because of the two filters we have used: a relatively high flux-density limit (which ensures that fractional observational uncertainties are low), and the rejection of sources well-described by a single-Gaussian (which ensures that the remaining sources are significantly variable). To test whether this is indeed the case we perform the same analysis, but for the fainter sources observed by OVRO that were not bright enough to be in our flux-density-limited subsample. For the fainter sources ($\mathrm{\langle S\rangle}\leq 0.354 ~Jy$, the flux-density limit)  the observational errors are relatively larger and will contribute more to the overall variance. We find that the fractional difference increases (Fig. \ref{plt:diff_comp_mod_noflux-limit}), with both tests (Wilcoxon, K-S) rejecting the null hypothesis that the two estimates of $\mathrm{\langle m\rangle}$ are drawn from the same distribution, as expected. Thus, for the brightest and most variable sources, the ability of the maximum likelihood approach to take errors into account does not provide a significant improvement on the overall estimate of $\mathrm{\langle m\rangle}$ over the simple direct standard deviation-over-mean estimate from the data. However, even for these sources the model introduced in this work has the advantages that: (a) it allows us to separate quiescent and flaring states; and (b) it provides information on the variability of each state, flaring ratio and duty cycle on a source-by-source basis.

\section{Variability properties}\label{Var_analysis}

\subsection{Overall variability properties of the sample}\label{over_var}
\begin{figure}
\resizebox{\hsize}{!}{\includegraphics[scale=1]{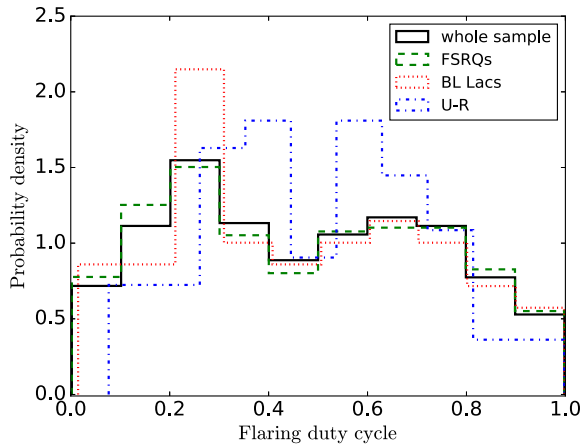} }
 \caption{Distribution of the fraction of time spent on the on-state (duty cycle) for the different populations (BL Lacs, FSRQs, U-Rs).}
 \label{plt:fraction_of_time}
 \end{figure}
 \begin{figure}
\resizebox{\hsize}{!}{\includegraphics[scale=1]{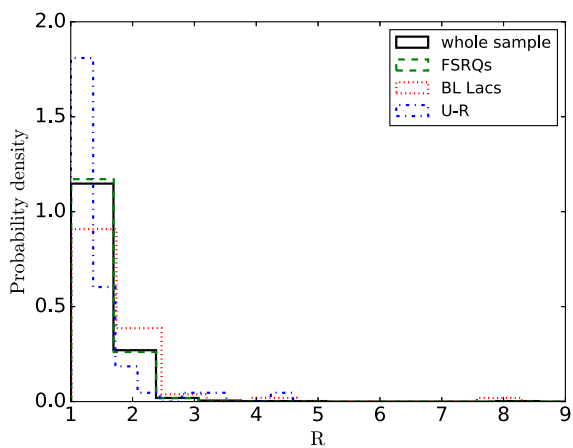} }
 \caption{Distribution of the flaring ratio $R$ for the different classes (BL Lacs, FSRQs, U-Rs).}
 \label{plt:ratio_on-off}
 \end{figure}

\begin{figure}
\resizebox{\hsize}{!}{\includegraphics[scale=1]{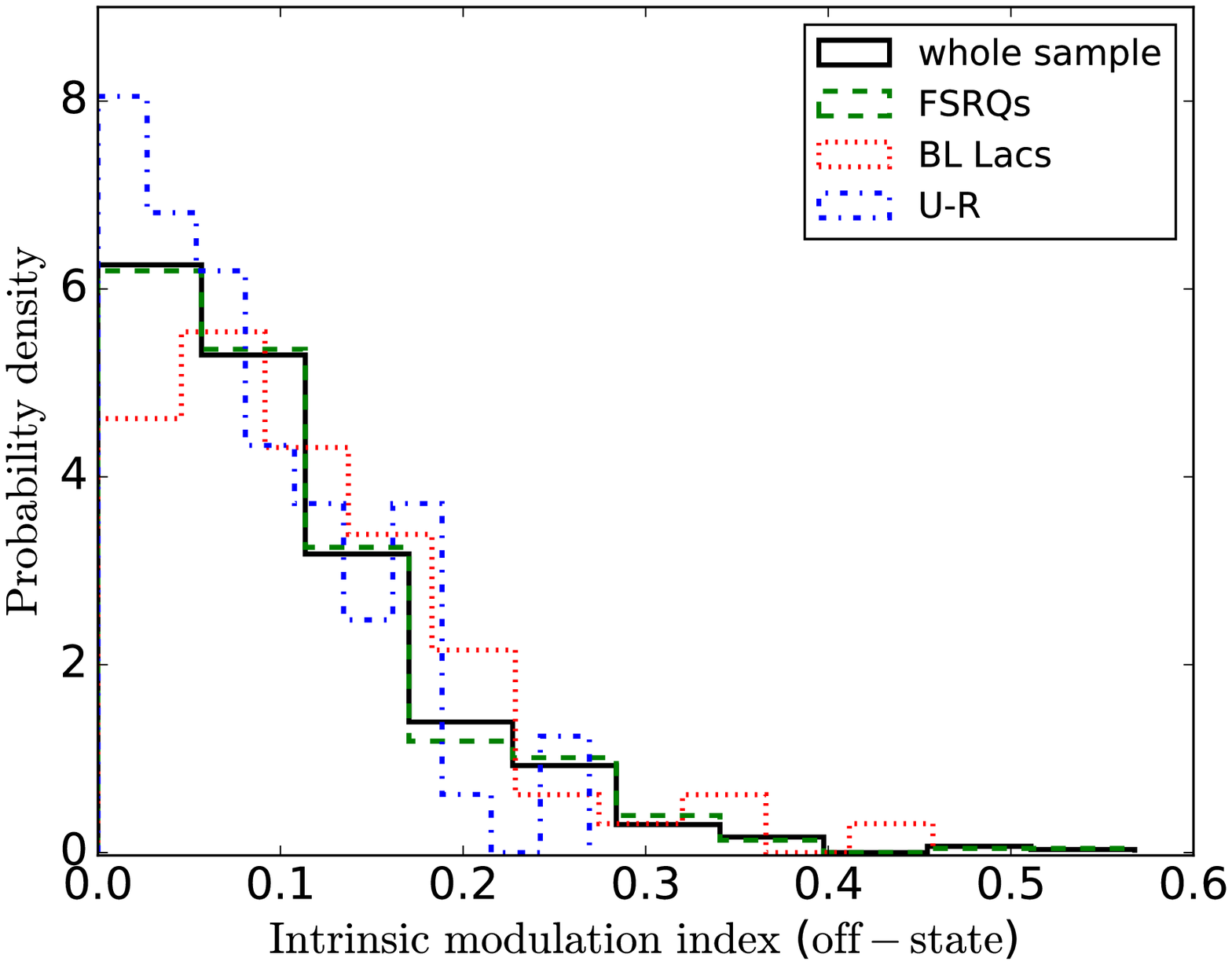} }
 \caption{Distribution of the intrinsic modulation index for the off-state for the different populations (BL Lacs, FSRQs, U-Rs).}
 \label{plt:modulation_off-state}
 \end{figure}
 \begin{figure}
\resizebox{\hsize}{!}{\includegraphics[scale=1]{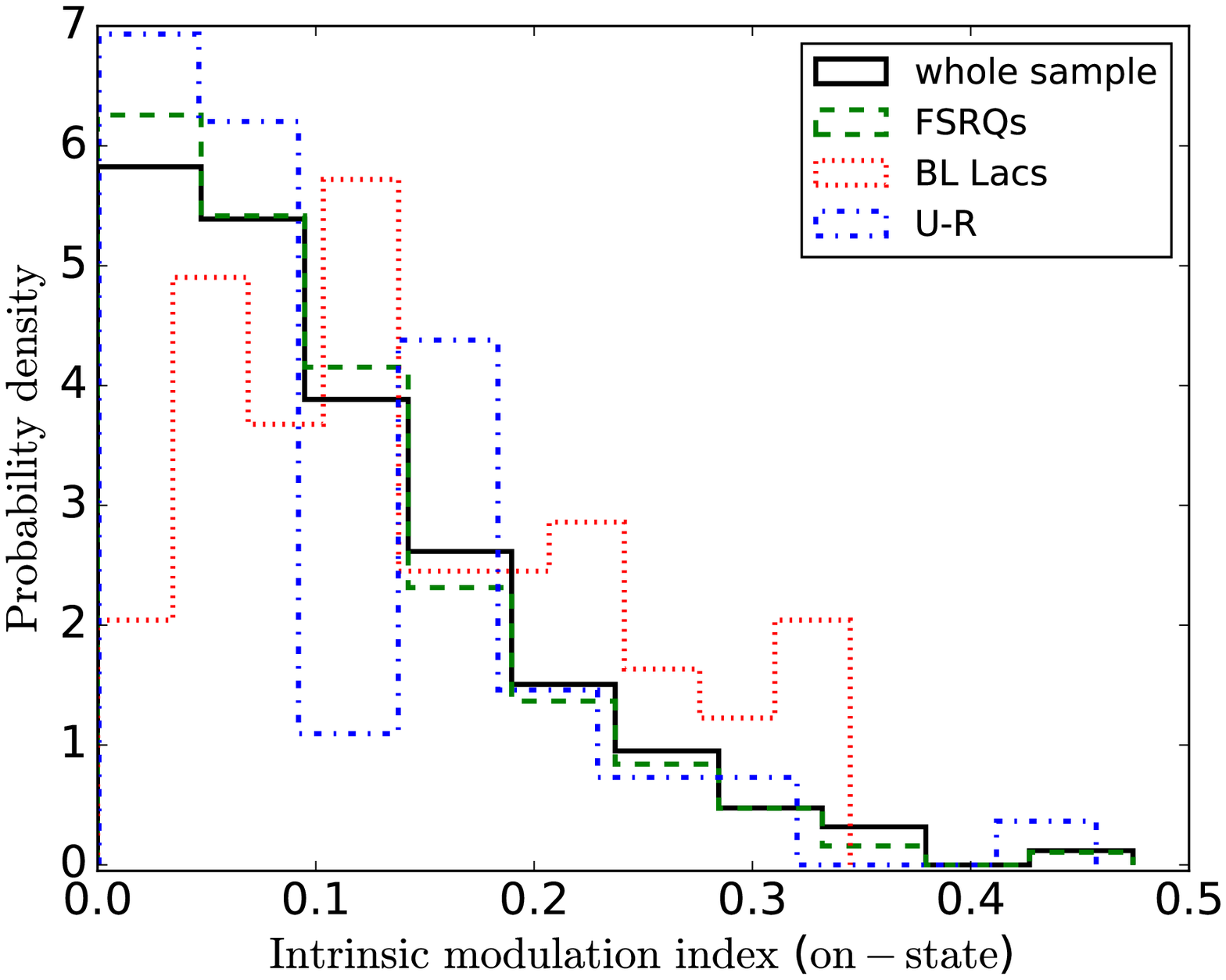} }
 \caption{Distribution of the intrinsic modulation index for the on-state for the different populations (BL Lacs, FSRQs, U-Rs).}
 \label{plt:modulation_on-state}
 \end{figure}

\begin{table}
\setlength{\tabcolsep}{12pt}
\centering
  \caption{Mean values for the duty cucle, flaring ratio and ``off''- and ``on''-state modulation indices for the entire sample}
  \label{tab:mean_sample}
\begin{tabular}{@{}ccccc@{}}
 \hline
 Parameter  & $f_{\rm t}$ & $R$ & $m_{\rm off}$ & $m_{\rm on}$\\
\hline 
  Mean & 0.47 &1.56 &0.10 & 0.10 \\
  Standard deviation & 0.26 & 1.55 & 0.08 & 0.08 \\
  \hline
\end{tabular}
\end{table}

\begin{table}
\setlength{\tabcolsep}{12pt}
\centering
  \caption{MLE exponential distributions for $R$, $m_{\rm on}$, $m_{\rm off}$, for the entire flux-density limited sample.}
  \label{tab:MLE_expo}
\begin{tabular}{@{}ccc@{}}
 \hline
    Quantity  & Mean & error on mean \\
   &  &     \\
  \hline

    $R$ &  1.485 & $\pm$ 0.006  \\
     $m_{\rm off}$ & 0.080  & $\pm$ 0.004 \\
    $m_{\rm on}$  & 0.089 & $\pm$ 0.004 \\
\hline
\end{tabular}
\end{table}

The distribution of the flaring duty cycle $f_{\rm t}$, the flaring ratio $R$, the intrinsic modulation index in the ``off''-state, $m_{\rm off}$ and the intrinsic modulation index in the ``on''-state, $m_{\rm on}$ for the entire sample, are shown in Figs. \ref{plt:fraction_of_time}, \ref{plt:ratio_on-off}, \ref{plt:modulation_off-state}, and \ref{plt:modulation_on-state} respectively with the black solid line. Table \ref{tab:mean_sample} shows the sample mean and standard deviation for these quantities.

It has been shown in \cite{Richards2011} that intrinsic modulation index follows a mono-parametric exponential distribution. This appears to be true for the individual ``on''- and ``off''- state modulation indices as well as the flaring ratio, while the duty cycle appears to be consistent with a uniform distribution for the flux-limited subsample. Parameterizing such quantities could prove a useful tool in modeling blazars at the population level. 

For the duty cycle, we test with the use of the K-S and Wilcoxon tests whether it is consistent with a uniform distribution in the [0,1] interval. Both tests cannot reject the null hypothesis that the duty cycle and a uniform distribution in the [0,1] range, are drawn from the same distribution (14\% and 19\% probability of consistency respectively). For this test, we have not excluded any sources based on their consistency with a single-Gaussian model. 

For the flaring ratio and modulation indices, the mean of the exponential distribution ($K_{\rm 0}$, where $K$ is the parameter to be fitted) given a set of observations ($K_{\rm obs,i}$) with Gaussian uncertainty $\sigma_{\rm obs,i}$ , can be calculated using,
\begin{eqnarray}
l_i &=&\frac{1}{2K_0}\exp\left[-\frac{K_{obs,i}}{K_0}\left(1-\frac{\sigma_{\rm obs,i}^2}{2K_0 K_{\rm obs,i}}\right) \right]\nonumber\\
&\times & \left\lbrace 1+\erf\left[\frac{K_{\rm obs,i}}{\sigma_{\rm obs,i}\sqrt{2}}\left(1-\frac{\sigma_{\rm obs,i}^2}{K_0K_{\rm obs,i}}\right)\right]\right\rbrace,
\label{eq:MLE_expo}
\end{eqnarray}
from \cite{Richards2011}, where $l_{\rm i}$ is the likelihood and erf is the error function. The error on the quantities to be fitted is calculated through integration of the normalized likelihood. For the modulation indices we exclude all values of $m<0.06$ and re-normalize the likelihood accordingly (see Eq. 30 and the corresponding discussion in \citealp{Richards2011}). 

Table \ref{tab:MLE_expo} shows parameters of the best-fit distributions for each of the aforementioned quantities which can be used as inputs to population models for blazars that require some treatment of source variability and/or activity state, such as the derivation of luminosity functions from single-epoch surveys. From the fits we have excluded sources J1433$-$1548, J1823+7938, J1808+4542, J1852+4019 for being outliers (either of $R$ or $m$), preventing the maximum likelihood method to achieve a good fit. In subsequent sections we focus on the comparison of variability properties between blazar subclasses.

\subsection{FSRQs versus BL Lacs}\label{QS_vs_BL}
\begin{figure}
\resizebox{\hsize}{!}{\includegraphics[scale=1]{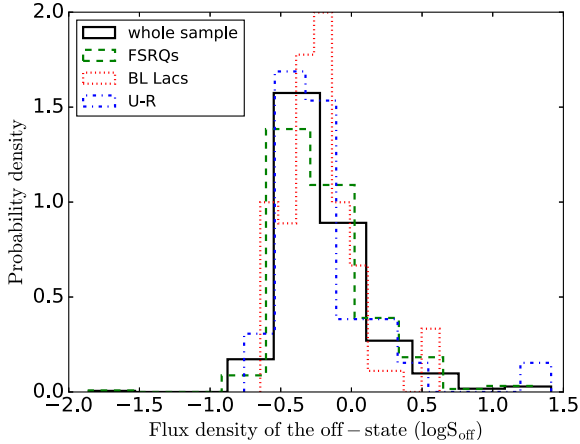} }
 \caption{Distribution of the off-state flux-densities for the different populations (BL Lacs, FSRQs, U-Rs).}
 \label{plt:flux_off-state}
 \end{figure}
 \begin{figure}
\resizebox{\hsize}{!}{\includegraphics[scale=1]{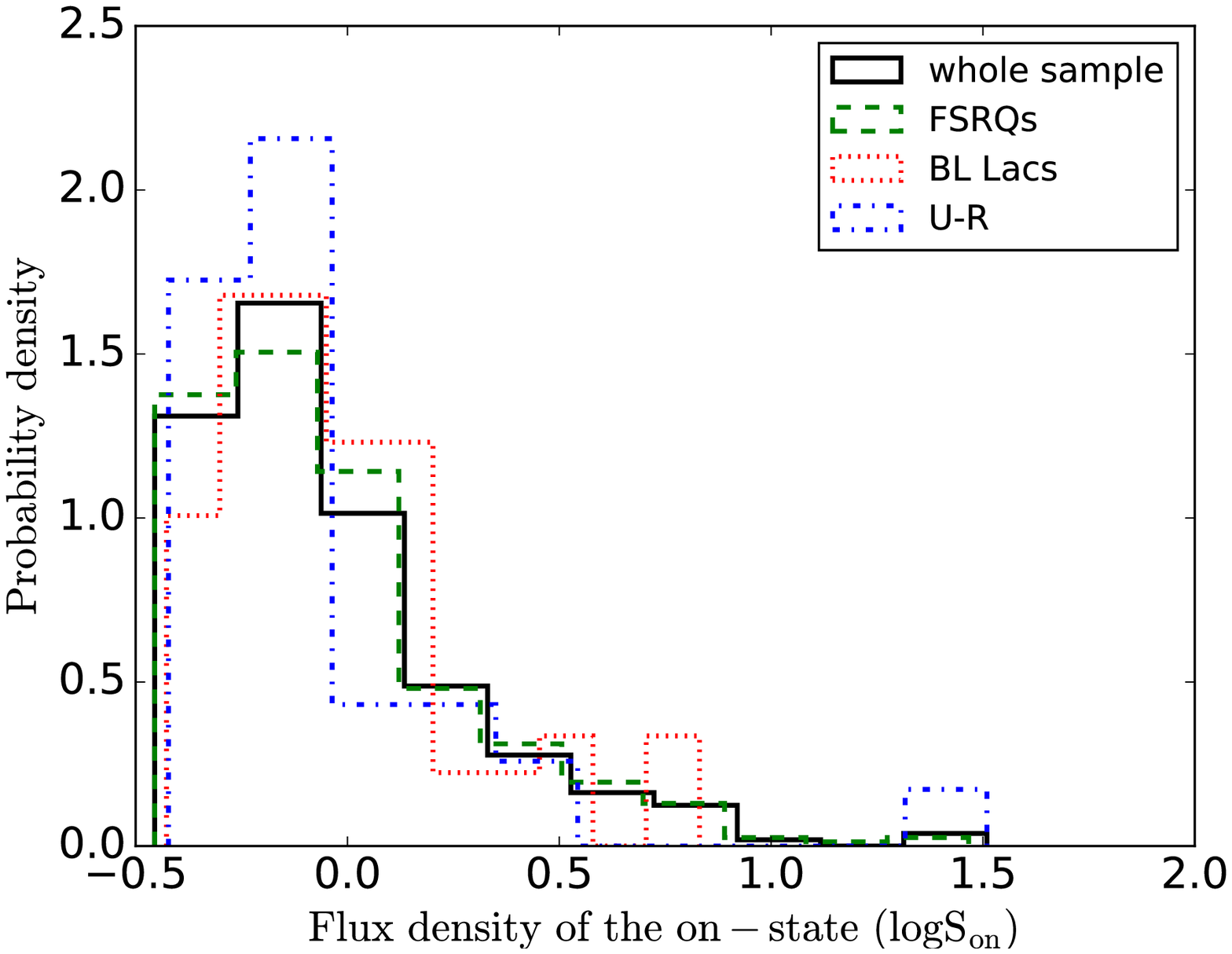} }
 \caption{Distribution of the on-state flux-densities for the different populations (BL Lacs, FSRQs, U-Rs).}
 \label{plt:flux_on-state}
 \end{figure}

\begin{table}
\setlength{\tabcolsep}{12pt}
\centering
  \caption{Mean parameters and Wilcoxon and K-S test probability values that the two population are consistent.}
  \label{tab:BL_vs_QS}
\begin{tabular}{@{}ccccc@{}}
 \hline
    Parameter  & BL Lacs & FSRQs & Wilcoxon & K-S \\
   & (mean) & (mean)  & (\%) & (\%)  \\
  \hline
    $f_{\rm t}$ & 0.46  & 0.47 & 92.3  & 86.1  \\
    $R$ &  1.71  & 1.55 & 0.3 & 0.8 \\
     $m_{\rm off}$& 0.11  & 0.10  & 32.8 & 64.6 \\
    $m_{\rm on}$  & 0.14  & 0.10 & $10^{-3}$ & 0.2 \\
\hline
\end{tabular}
\end{table}

Figure \ref{plt:fraction_of_time} shows the distribution of the fraction of time spent on the ``on'' state (the blazar flaring duty cycle) for the whole sample (black solid line), BL Lacs (red dotted line), FSRQs (green dashed line), and U-Rs (blue dashed-dotted line). The distribution is similar to a uniform one with no apparent difference between BL Lacs and FSRQs. This is also confirmed by the K-S test and the Wilcoxon test.

We next explore the flaring ratio of the flux-density of the ``off'' and``on'' states  which is unaffected by redshift effects. This way we are not limited by the redshift completeness of our sample. 

Figure \ref{plt:ratio_on-off} shows the ratio of the ``on''-to-``off''-state ($R$) for the different populations. The BL Lacs appear to have larger $R$ (on average) than the FSRQs (Table \ref{tab:BL_vs_QS}). Both tests reject the null hypothesis that the samples are drawn from the same distribution with the values of one (BL Lacs) being systematically larger than the other. This would suggest that while flaring the BL Lacs reach, on average, higher flux-densities on the on-state relatively to their off-states than the FSRQs.

Although the ratio $R$ is different for the BL Lacs and FSRQs, the individual flux-densities of the ``off'' (Fig. \ref{plt:flux_off-state}) and ``on'' states (Fig. \ref{plt:flux_on-state}) appear to be similar with the K-S ($45\%$ for the ``off'' and $46\%$ for the ``on'' state) and Wilcoxon ($83\%$ for the off and $32\%$ for the on-state) tests unable to reject the null hypothesis, suggesting that the BL Lacs exhibit relatively larger outbursts than the FSRQs.

We also explore the characteristic variability of the populations using the intrinsic modulation index defined as $m=\sigma/S$ \citep{Richards2011}. Figures \ref{plt:modulation_off-state} and \ref{plt:modulation_on-state} show the distribution of $m$ for the ``off'' and ``on'' states respectively.

For the ``off''-state the two tests cannot reject the null hypothesis that the two samples are drawn from the same distribution. However, for the ``on''-state the mean for the BL Lacs is significantly larger than the FSRQs both tests rejecting the hypothesis that the two samples are drawn from the same distribution. The fact that the mean ``on''-state modulation index value is larger for the BL Lacs suggests that while flaring their flux-density distribution is on average wider. On the other hand, for the FSRQs, there is no difference of the intrinsic modulation index between the two states. We conclude that the BL Lacs are relatively more variable during outbursts than the FSRQs. All the mean parameter and probability values are summarized in Table  \ref{tab:BL_vs_QS}.

\subsection{Blazars versus Unidentified}\label{BZ_vs_UR}
\begin{table}
\setlength{\tabcolsep}{12pt}
\centering
  \caption{Mean parameters and Wilcoxon and K-S test probability values that the two populations are consistent.}
  \label{tab:BZ_vs_UR}
\begin{tabular}{@{}ccccc@{}}
 \hline
    Parameter  & Blazars & U-Rs & Wilcoxon & K-S \\
   & (mean) & (mean)  & (\%) & (\%)  \\
  \hline
    $f_t$ & 0.47  & 0.50 & 32.9  & 13.3  \\
    $R$ & 1.57  & 1.44 & 0.08 & 0.01 \\
     $m_{\rm off}$&  0.10 & 0.08 & 9.0 & 30.8 \\
    $m_{\rm on}$  & 0.11  & 0.10 & 46.7 & 33.1 \\
\hline
\end{tabular}
\end{table}

\begin{table}
\setlength{\tabcolsep}{12pt}
\centering
  \caption{Mean parameters and Wilcoxon and K-S test probability values that the two populations are consistent.}
  \label{tab:BL_vs_UR}
\begin{tabular}{@{}ccccc@{}}
 \hline
    Parameter  & BL Lacs & U-Rs & Wilcoxon & K-S \\
   & (mean) & (mean)  & (\%) & (\%)  \\
  \hline
    $f_{\rm t}$ & 0.46  & 0.50 & 34.4  & 33.9  \\
    $R$ & 1.71  & 1.44 & 0.007 & 0.009 \\
     $m_{\rm off}$&  0.11 & 0.08 & 0.3 & 0.4 \\
    $m_{\rm on}$  & 0.14  & 0.10 & 5.9 & 27.0 \\
\hline
\end{tabular}
\end{table}

Comparing the blazar sample (FSRQs + BL Lacs) against the U-R sources we find very interesting similarities. The duty cycle is similar, with neither test rejecting the null hypothesis that the samples are drawn from the same distribution (Table \ref{tab:BZ_vs_UR}). The mean flaring ratio $R$ is mildly (but significantly) larger for the blazars than for the U-Rs, in both states the blazars have larger mean modulation index, yet both tests are unable to reject the null hypothesis that the two samples are drawn from the same distribution (Table \ref{tab:BZ_vs_UR}).

 We also compare individual classes (i.e., FSRQs , BL Lacs) versus U-Rs. Comparing the FSRQs none of the above results changes in any significant manner, which is to be expected since the FSRQs dominate the blazar sample (75\% of the sample). For the BL Lacs the results are somewhat different. The duty cycle as well as the ``off''-state modulation index are similar, with both tests not being able to reject the null hypothesis that the two samples are drawn from the same distribution. However, both tests reject the null hypothesis for the flaring ratio $R$  and for the ``on''-state modulation index (Table \ref{tab:BL_vs_UR}).

If the unidentified and blazar candidate sources in the U-R sample were unidentified BL Lacs due to the absence of spectral lines, we would expect the opposite results, i.e., the BL Lac sample to be more consistent with the U-R sample in the variability characteristics. Instead the FSRQs appear to be consistent with the U-Rs suggesting that the majority of the sources in the U-R sample are either unidentified FSRQs or radio galaxies with jets pointed close to our line of sight, yet not close enough to be considered a blazar (often referred to as misaligned blazars). The latter would suggest that the FSRQs and the radio galaxies share similar variability characteristics. All the mean parameter and probability values are summarized in Table  \ref{tab:BZ_vs_UR}.

\subsection{{\it Fermi}-detected versus {\it Fermi} non-detected blazars}\label{GL_vs_GQ}
\begin{table}
\setlength{\tabcolsep}{12pt}
\centering
  \caption{Mean parameters and Wilcoxon and K-S test probability values that the two populations are consistent.}
  \label{tab:Fermi_vs_noFermi}
\begin{tabular}{@{}ccccc@{}}
 \hline
    Parameter  & {\it Fermi} & non-{\it Fermi} & Wilcoxon & K-S \\
   & (mean) & (mean)  & (\%) & (\%)  \\
  \hline
    $f_{\rm t}$ & 0.47  & 0.48 & 65.4  & 52.8  \\
    $R$ &  1.60  & 1.51 & $10^{-10}$ & $10^{-8}$  \\
     $m_{\rm off}$& 0.11  & 0.09 & 0.009 & 0.26 \\
    $m_{\rm on}$  & 0.13  & 0.09 & $10^{-7}$ & 0.001 \\
\hline
\end{tabular}
\end{table}
\begin{figure}
\resizebox{\hsize}{!}{\includegraphics[scale=1]{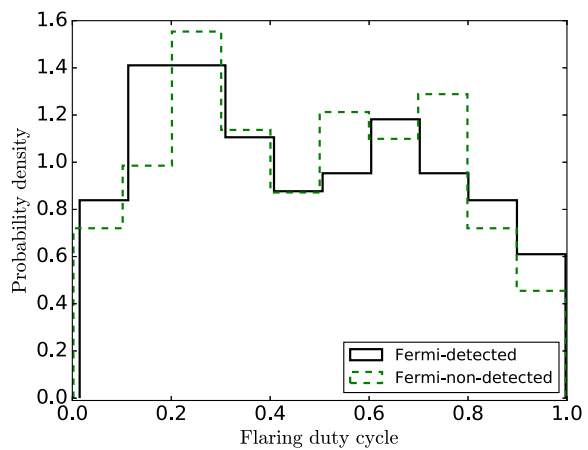} }
 \caption{Distribution of the fraction of time spent on the on-state (duty cycle) for the {\it Fermi}-detected and non-detected sources.}
 \label{plt:fraction_of_time_fermi}
 \end{figure}
\begin{figure}
\resizebox{\hsize}{!}{\includegraphics[scale=1]{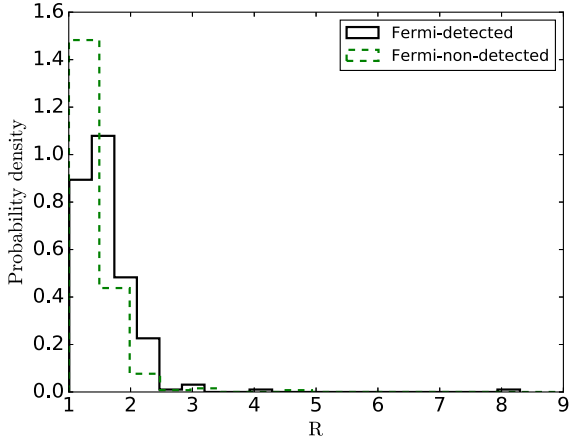} }
 \caption{Distribution of the flaring ratio $R$ for the for the {\it Fermi}-detected and non-detected sources.}
 \label{plt:ratio_on-off_fermi}
 \end{figure}
\begin{figure}
\resizebox{\hsize}{!}{\includegraphics[scale=1]{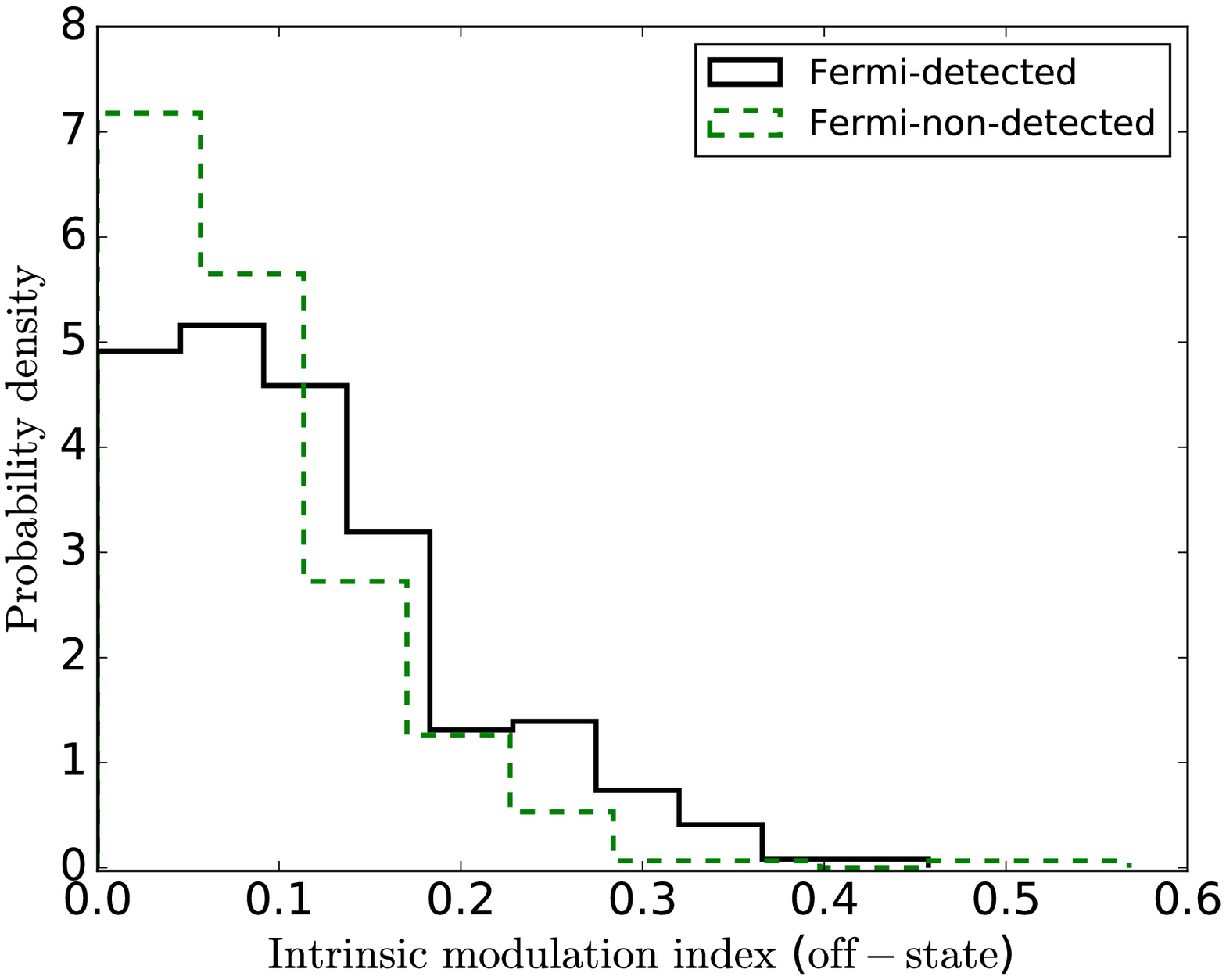} }
 \caption{Distribution of the intrinsic modulation index for the off-state for the {\it Fermi}-detected and non-detected sources.}
 \label{plt:modulation_off-state_fermi}
 \end{figure}
 \begin{figure}
\resizebox{\hsize}{!}{\includegraphics[scale=1]{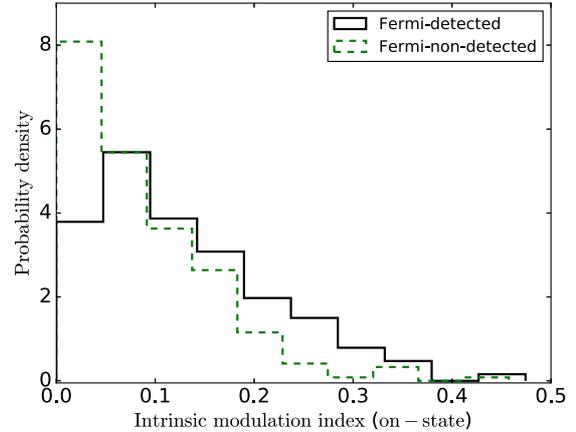} }
 \caption{Distribution of the intrinsic modulation index for the on-state for the {\it Fermi}-detected and non-detected sources.}
 \label{plt:modulation_on-state_fermi}
 \end{figure}
There are 267 sources in our flux-density-limited sample (50\%) that have been associated with {\it Fermi} detected sources. Out of the 267 sources, 186 are FSRQs, 62 are BL Lacs, and 19 are U-Rs. Figure \ref{plt:fraction_of_time_fermi} shows the distribution of the duty cycle. The solid black line is for the {\it  Fermi}-detected and the dashed green for the {\it Fermi}-non-detected sources. The distributions are very similar in shape with the Wilcoxon and K-S test suggesting the two samples being drawn from the same distribution (Table \ref{tab:Fermi_vs_noFermi}). The high probability of consistency, as well as the almost identical mean values suggest that there is no difference in the time the two samples spent in either state.

However, their flaring ratio $R$ (Fig. \ref{plt:ratio_on-off_fermi}) is rather different. Both tests rejected the null hypothesis of consistency between populations with high confidence (i.e., small probability value). The mean values combined with the results of the Wilcoxon test suggest that the {\it Fermi}-detected sources have systematically higher ratios, which translates to them having larger differences in flux-density between states, i.e., more powerful outbursts.

Examining the intrinsic modulation index in the different states (Fig. \ref{plt:modulation_off-state_fermi}, \ref{plt:modulation_on-state_fermi}), we find that in both the ``off'' and the ``on'' state, both tests rejected the null hypothesis. In both states the {\it Fermi}-detected sources have larger mean values which means that on average they are more variable than the {\it Fermi}-non-detected sources. At the same time there is no difference in the variability of the two states in the {\it Fermi}-non-detected sources, contrary to the {\it Fermi}-detected.  This could partially be attributed to the fact that the majority of the BL Lacs (which are more variable in the ``on'' state, see section \ref{QS_vs_BL}) are in the {\it Fermi}-detected subsample, however, the fact that the {\it Fermi}-detected are systematically more variable than the {\it Fermi}-non-detected in both states, is in support of the {\it Fermi}-detected being intrinsically more variable than the {\it Fermi}-not-detected sources (see also \cite{Richards2014}). All the mean parameter and probability values are summarized in Table  \ref{tab:Fermi_vs_noFermi}.

\subsection{Low versus High redshift sources}\label{low_vs_high}

\cite{Richards2011} and  \cite{Richards2014} point out that FSRQs show a negative correlation between radio variability as quantified by their intrinsic modulation index and redshift ($z$). They find that sources with $z<1$ are more variable in radio than sources with $z>1$ suggesting that the radio variability of FSRQs evolves with cosmic time. The trend first appeared in the two year dataset \citep{Richards2011}, and persisted in the four year dataset \citep{Richards2014}. We use the redshift values from \cite{Richards2014} to separate our sources into two subsamples according to their redshift. All the FSRQs in our sample have a known redshift. Separating our samples in low ($z<1$) and high ($z>1$) redshift sources we have 149 and 252 sources respectively.

 For the duty cycle and the intrinsic modulation index in the ``on''-state, both tests agree that the two subsamples are drawn from the same distributions. For the  flaring ratio and the ``off''-state modulation index the tests reject the null hypothesis. If instead of $z=1$ we separate our samples according to the mean ($z\approx 1.22$) or median ($z\approx 1.32$) value the results of both test vary for all parameters. Sensitivity to redshift separation makes the individual components of the analysis ($f_{\rm t}$, $R$, $m_{\rm off}$, $m_{\rm on}$) unreliable indicators for cosmic evolution. However, if we compare the overall modulation index ($\mathrm{\langle m\rangle}$) we find that regardless of redshift separation the two samples are inconsistent. Separating the samples at $z=1$ and $z=\langle z \rangle_{\rm median}$ the probability values of both tests are $\leq 10^{-3}$. If we separate the samples at $z=\langle z\rangle_{\rm mean}$ the Wilcoxon test yields a $2\%$, whereas the K-S test yields a $6\%$ probability of consistency which is marginally acceptable. Since the mean $\langle m\rangle$ values for the two samples are different ($\langle m\rangle_{\rm mean}=0.185$ for the low and $\langle m\rangle_{\rm mean}=0.177$ for the high redshift sources) we conclude that the two samples are not drawn from the same distribution. On the other hand, if we correct for the cosmological time dilation ($\Delta t_{\rm obs}=(1+z)\Delta t_{\rm rest-frame}$) and repeat the analysis, but this time using equal redshift-corrected observing lengths for all sources, we find no apparent trend of cosmic evolution. For all the parameters both tests cannot reject the null hypothesis that the two samples are drawn from the same distribution. However, we have not taken into account relativistic compression of variability timescales, which should be larger for higher-z sources (because of flux-density selection effects) and thus affect results in the opposite direction to cosmological time dilation. Given the large span of Doppler factors in blazar jets \citep{Hovatta2009,Liodakis2017-II} even the redshift-corrected observing lengths will be significantly different from the ``true'' jet rest-frame observing lengths for each source. Until a large enough number of Doppler factor estimates is available that will allow us to confidently correct for the relativistic effects on a source-by-source basis, we are unable to drawn firm conclusions on the cosmic evolution of FSRQ variability properties. 
 
A similar effect as the one discussed here and in \cite{Richards2011,Richards2014} is true for interstellar scintillation at 5~GHz  \citep{Lovell2008,Koay2012}. The authors found that interstellar scintillation is suppressed a higher redshift and that high redshift sources have steeper spectra (in the 5-9~GHz range). That the interstellar scintillation of high redshift sources is suppressed indicates either a larger apparent angular size, beyond the expected cosmological expansion, or a smaller compact fraction in the high redshift blazars. However, since the OVRO dataset is single frequency we cannot test if a similar effect is present in our sample.

\section{Summary}\label{discussion}

We have presented a novel five-dimensional maximum likelihood formalism in order to characterize the variability properties of blazars and blazar-like sources as a series of two states: an ``off'' state describing the low activity periods of a source, and an ``on'' state that describes periods of outbursts. We used our method to fit the 15 GHz flux-density distribution of blazars as seen by OVRO \citep{Richards2011}, and extract their variability properties (Table \ref{online_data}).

For our statistical analysis of the population properties of the OVRO blazars, as well as for the comparison of the behaviors of blazar subclasses, we define a statistically complete subset of all OVRO monitored sources: a flux-density limited subsample, based on average fluxes from \cite{Richards2014}. We have tested whether the model assumed in this work is a good description of the underlying flux-density distribution of the OVRO-monitored blazars. If this is the case, then for bright sources (where observational uncertainties do not widen appreciably the observed flux density distribution compared to the intrinsic one) we would expect the value of $m$ obtained from the sample mean and sample standard deviation for each light curve to agree well with our likelihood-derived intrinsic $\mathrm{\langle m\rangle}$. We have verified that the two are, in fact, in excellent agreement.

Having established that our model is a good description of the underlying distribution, we compared the variability characteristics of different subsamples. Our results can be summarized as follows:

\begin{itemize}

{\item BL Lacs are more variable than the FSRQs. This is consistent with the finding of \cite{Richards2011,Richards2014} that BL Lacs have a higher overall $\mathrm{\langle m\rangle}$. However, we have now established that BL Lacs also exhibit stronger outbursts (have a higher flaring ratio), and that their increase in $\mathrm{\langle m\rangle}$ is dominated by the ``on''-state (it is $m_{\rm on}$ that is significantly higher, while $m_{\rm off}$ is similar in BL Lacs and FSRQs). Interestingly, \cite{Liodakis2015,Liodakis2017-II} find that the Doppler factors of FSRQs are on average, significantly higher than those of BL Lacs, so this discrepancy must have its origin in rest-frame properties, rather than differences in boosting between the two classes.}

{\item Sources classified as blazars (BL Lacs and FSRQs) have systematically larger flaring ratios (i.e., stronger outbursts) than the U-Rs with otherwise similar variability characteristics. The variability characteristics of U-Rs are similar to FSRQs. Since the majority of U-Rs are blazar candidates, it would suggest either that they are unclassified FSRQs most likely due to lack of multi-wavelength observations, or that they are unclassified radio galaxies, which in turn would suggest that FSRQs and radio galaxies share similar rest-frame variability characteristics.
}

{\item {\it Fermi}-detected sources are intrinsically more variable than the {\it Fermi}-non-detected sources. This result agrees with the overall findings of \cite{Richards2011,Richards2014}, however the bimodal model offers an opportunity to trace the origin of this result in the details of the behavior of blazars in the flux density domain. Indeed, {\it Fermi}-detected sources have higher flaring ratios and higher modulation indices in both states, with the most significant difference being in the flaring ratio. This results indicates that the mechanisms responsible for the amplitude of radio variations and the $\gamma$-ray loudness of a source may share a physical link.

}

{\item The overall intrinsic modulation index ($\langle m\rangle$) is consistently (regardless of redshift separation of subsamples) supporting the negative correlation between radio variability and redshift in FSRQs reported in \cite{Richards2011,Richards2014}. Once we accounted for the cosmological time dilation we found no evidence for such negative correlation. However, since we are not yet able to take into account the relativistic effects compressing blazar timescales, we are not able to come to firm conclusions regarding a possible cosmic evolution of variability properties of FSRQs.

}

\end{itemize}

We caution the reader that for an analysis such as the one presented here, there is a dependence of the derived variability parameters to the length of the monitoring program. Short (in time) monitoring programs may not be able to sample the entire flux-density distribution of a blazar. However, given enough time, estimates will converge to their ``true'' values. For our well sampled sources, we find that by splitting the monitoring period in half (4 years) the difference in the derived variability estimates from the two periods is $<50\%$ in the majority of cases.

The overall radio modulation index as calculated by \cite{Richards2011} is one of the properties used to select samples for monitoring of other blazar properties (such as their optopolarimetric behavior e.g., \citealp{King2014,Pavlidou2014}). Given that we find the bimodal flux distribution to be a much better description than the single Gaussian used by \cite{Richards2011} we would advice using the overall  $\mathrm{\langle m\rangle}$ values from this work for sample characterization.

The tools presented in this work were used to explore the variability properties of a 15 GHz selected flux-density-limited sample. They are, however, not restricted to any particular frequency since the formalism is based on statistics alone and is independent of any emission mechanism or other physical arguments. Thus, are suitable for all wavelengths and sources that can be well described by a bimodal Gaussian distribution. However, one should keep in mind that using a large number of observations for the fitting (in our case 421 on average) is critical to ensure robust results for the estimated parameters and their errors.

\section*{Acknowledgments}
The authors would like to thank the referee Hayley Bignall and Vassilis Karamanavis for constructive comments and suggestions that helped improve this work. I.L. thanks the Caltech Astronomy Department for their hospitality during the completion of this work. This research was supported by the European Commission Seventh Framework Program (FP7) through grants PCIG10-GA-2011-304001 ``JetPop'' and PIRSES-GA-2012-31578 ``EuroCal''. T.H. was supported in part by the Academy of Finland project number 267324. This research has made use of data from the OVRO 40-m monitoring program \citep{Richards2011} which is supported in part by NASA grants NNX08AW31G, NNX11A043G, and NNX14AQ89G and NSF grants AST-0808050 and AST-1109911.

\bibliographystyle{mnras}
\bibliography{bibliography} 

\pagestyle{empty}
\begin{landscape}
\begin{table}
\centering
\setlength{\tabcolsep}{2.7pt}
\caption{Variability parameters for the OVRO monitored sources. Columns: (1) OVRO name; (2) duty cycle ($f_{\rm t}$); (3) 1$\sigma$ error on $f_{\rm t}$ from the information matrix; (4) upper 1$\sigma$ error on $f_{\rm t}$ from the likelihood slice; (5) lower 1$\sigma$ error on $f_{\rm t}$ from the likelihood slice; (6) off-state flux-density ($S_{\rm off}$); (7) 1$\sigma$ error on $S_{\rm off}$ from the information matrix; (8) upper 1$\sigma$ error on $S_{\rm off}$ from the likelihood slice; (9) lower 1$\sigma$ error on $S_{\rm off}$ from the likelihood slice; (10) standard deviation of the off-state flux-density ($\sigma_{\rm off}$); (11) 1$\sigma$ error on $\sigma_{\rm off}$ from the information matrix; (12) upper 1$\sigma$ error on $\sigma_{\rm off}$ from the likelihood slice; (13) lower 1$\sigma$ error on $\sigma_{\rm off}$ from the likelihood slice; (14) on-state flux-density ($S_{\rm on}$); (15) 1$\sigma$ error on $S_{\rm on}$ from the information matrix; (16) upper 1$\sigma$ error on $S_{\rm on}$ from the likelihood slice; (17) lower 1$\sigma$ error on $S_{\rm on}$ from the likelihood slice; (18) standard deviation of the on-state flux-density ($\sigma_{\rm on}$); (19) 1$\sigma$ error on $\sigma_{\rm on}$ from the information matrix; (20) upper 1$\sigma$ error on $\sigma_{\rm on}$ from the likelihood slice; (21) lower 1$\sigma$ error on $\sigma_{\rm on}$ from the likelihood slice. The parameter values for $S_{\rm off}$, $\sigma_{\rm off}$, $S_{\rm on}$ and $\sigma_{\rm on}$ are in Jansky. For uncertainty values in columns (3), (7), (11), (15), and (19) equal to zero, the information matrix failed to provide an error estimate. In this case, use upper and lower 1-sigma uncertainties from the likelihood slice. The full-version of the table is available online.}
\label{online_data}
\tiny
\begin{tabular}{@{}ccccccccccccccccccccc}
\hline
Name & $f_{\rm t}$ & $\sigma_{\rm f_t}$ & upper-$\sigma_{f_{\rm t}}$ & lower-$\sigma_{f_{\rm t}}$ & $S_{\rm off}$ & $\sigma_{S_{\rm off}}$ & upper-$\sigma_{S_{\rm off}}$  & lower-$\sigma_{S_{\rm off}}$ & $\sigma_{\rm off}$ & $\sigma_{\sigma_{\rm off}}$ & upper-$\sigma_{\sigma_{\rm off}}$  & lower-$\sigma_{\sigma_{\rm off}}$  & $S_{\rm on}$ & $\sigma_{S_{\rm on}}$ & upper-$\sigma_{S_{\rm on}}$  & lower-$\sigma_{S_{\rm on}}$ & $\sigma_{\rm on}$ & $\sigma_{\sigma_{\rm on}}$ & upper-$\sigma_{\sigma_{\rm on}}$  & lower-$\sigma_{\sigma_{\rm on}}$ \\
(1) &(2) &(3) &(4) &(5) &(6) &(7) &(8) &(9) &(10) &(11) &(12) &(13) &(14) &(15) &(16) &(17) &(18) &(19) &(20) &(21) \\
\hline
J0001-1551 & 0.168 & 0.0373 & + 0.0369 & - 0.0369 & 0.201 & 0.0018 & + 0.002 & - 0.002 & 0.0 & 0.0048 & + 0.0037 & - 0.0 & 0.222 & 0.0 & + 0.0044 & - 0.0044 & 0.044 & 0.0 & + 0.0022 & - 0.0022 \\ 
J0001+1914 & 0.839 & 0.0208 & + 0.0252 & - 0.0252 & 0.229 & 0.0017 & + 0.0023 & - 0.0023 & 0.027 & 0.0014 & + 0.0016 & - 0.0013 & 0.311 & 0.0 & + 0.0031 & - 0.0031 & 0.009 & 0.0 & + 0.0023 & - 0.0021 \\
J0003+2129 & 0.978 & 0.012 & + 0.0196 & - 0.0196 & 0.077 & 0.0006 & + 0.0008 & - 0.0008 & 0.008 & 0.0005 & + 0.0006 & - 0.0005 & 0.083 & 0.0 & + 0.0132 & - 0.0124 & 0.033 & 0.0 & + 0.0115 & - 0.0072 \\
J0004-1148 & 0.205 & 0.0237 & + 0.0224 & - 0.0216 & 0.366 & 0.0061 & + 0.0059 & - 0.0059 & 0.04 & 0.0046 & + 0.0045 & - 0.004 & 0.631 & 0.0 & + 0.0069 & - 0.0069 & 0.113 & 0.0 & + 0.0055 & - 0.0052 \\
J0004+2019 & 0.59 & 0.0303 & + 0.0277 & - 0.0283 & 0.289 & 0.0016 & + 0.0014 & - 0.0014 & 0.015 & 0.0015 & + 0.0014 & - 0.0013 & 0.374 & 0.0 & + 0.0049 & - 0.0049 & 0.049 & 0.0 & + 0.0034 & - 0.0032 \\
J0004+4615 & 0.068 & 0.0121 & + 0.0129 & - 0.0116 & 0.049 & 0.0021 & + 0.0025 & - 0.0025 & 0.0 & 0.0044 & + 0.0038 & - 0.0 & 0.207 & 0.018 & + 0.0041 & - 0.0041 & 0.041 & 0.0134 & + 0.0017 & - 0.0017 \\
J0005+0524 & 0.997 & 0.0026 & + 0.01 & - 0.01 & 0.107 & 0.0006 & + 0.0011 & - 0.0011 & 0.007 & 0.0005 & + 0.0006 & - 0.0006 & 0.172 & 0.0 & + 0.0103 & - 0.0103 & 0.0 & 0.0 & + 0.0124 & - 0.0 \\
J0005-1648 & 0.526 & 0.0801 & + 0.0421 & - 0.0421 & 0.131 & 0.0037 & + 0.0026 & - 0.0026 & 0.014 & 0.0024 & + 0.0014 & - 0.0013 & 0.159 & 0.0 & + 0.0016 & - 0.0016 & 0.008 & 0.0 & + 0.0013 & - 0.0012 \\
J0005+3820 & 0.839 & 0.0181 & + 0.0176 & - 0.0185 & 0.509 & 0.003 & + 0.0031 & - 0.0031 & 0.053 & 0.0024 & + 0.0024 & - 0.0023 & 0.698 & 0.0 & + 0.0049 & - 0.0049 & 0.029 & 0.0 & + 0.0049 & - 0.0042 \\
J0006-0623 & 0.795 & 0.023 & + 0.0223 & - 0.0239 & 2.243 & 0.0089 & + 0.0112 & - 0.0112 & 0.141 & 0.0068 & + 0.0075 & - 0.0071 & 3.413 & 0.0 & + 0.0819 & - 0.0785 & 0.572 & 0.0 & + 0.0595 & - 0.0526 \\
\hline
\end{tabular}
\end{table}
\end{landscape}
\clearpage   
\pagestyle{plain}

\label{lastpage}
\end{document}